\documentclass[italian,english]{article}
\usepackage[latin9]{inputenc}
\usepackage{amssymb}
\usepackage{esint}

\newcommand{\lyxaddress}[1]{
\par {\raggedright #1
\vspace{1.4em}
\noindent\par}
}

\usepackage{babel}

\begin{document}

\title{\textbf{Removing black hole singularities with nonlinear electrodynamics }}

\author{\textbf{$^{1}$Christian Corda and $^{2}$Herman J. Mosquera Cuesta}}

\maketitle

\lyxaddress{\begin{center}
\textbf{$^{1}$}Associazione Galileo Galilei, Via Pier Cironi 16
- 59100 Prato, Italy; \textbf{$^{2}$}Instituto de Cosmologia, Relatividade
e Astrofìsica (ICRA-BR), Centro Brasilero de Pesquisas Fisicas, Rua
Dr. Xavier Sigaud 150, CEP 22290 -180 Urca Rio de Janeiro - RJ Brazil 
\par\end{center}}

\begin{center}
\textit{E-mail addresses:} \textbf{$^{1}$}\emph{cordac.galilei@gmail.com;
}\textbf{$^{2}$}\emph{herman@icra.it}
\par\end{center}
\begin{abstract}
We propose a way to remove black hole singularities by using a particular
nonlinear electrodynamics Lagrangian that has been recently used in
various astrophysics and cosmological frameworks. In particular, we
adapt the cosmological analysis discussed in a previous work to the
black hole physics. Such analysis will be improved by applying the
Oppenheimer\textendash{}Volkoff equation to the black hole case. At
the end, fixed the radius of the star, the final density depends only
on the introduced \emph{quintessential density term} $\rho_{\gamma}$
and on the mass.

\medskip{}

\emph{Keywords}: Black holes, removed singularity.
\end{abstract}
It is well known that the concept of black hole has been considered
very fascinating by scientists even before the introduction of Einstein\textquoteright{}s
general relativity (see Ref. \cite{key-1} for a historical review).
However, an unsolved problem concerning such objects is the presence
of a spacetime singularity in their core. Such a problem was present
starting by the first historical papers concerning black holes \cite{key-2}-\cite{key-4}
and was generalized in the famous paper by Penrose \cite{key-5}.
It is a common opinion that this problem could be solved when a correct
quantum gravity theory will be, finally, constructed (see Ref. \cite{key-6}
for recent developments). 

In this letter, we propose a way to remove black hole singularities
at a classical level. The idea is to use a particular nonlinear electrodynamics
Lagrangian to address this issue. Such Lagrangian has been recently
used in various analyses in astrophysics, like surface of neutron
stars \cite{key-7} and pulsars \cite{key-8} and also on cosmological
contexts \cite{key-9}. In particular, we will adapt the cosmological
analysis in Ref. \cite{key-9} to the black hole case. The analysis
will be improved by applying the Oppenheimer\textendash{}Volkoff equation
\cite{key-15} to the black hole case. In this way, we show that the
density singularity of a black hole is also removed. At the end, fixed
the radius of the star, the final density depends only on $\rho_{\gamma}$,
a a \emph{quintessential density term} which will be introduced in
the following, and also on the star mass. 

At this point it is very important to provide to the reader some elements
that were part of our motivations for using NonLinear Electrodynamics
(NLED) in the search for physical methods to eliminate the black hole
singularity in general relativity \cite{key-19}. Those elements provide
a good guide to understand the formal analysis that we are going to
present below. 

As it has been carefully explained in Refs. \cite{key-7} and \cite{key-8},
the effects arising from a NLED become quite important in super-strongly
magnetized compact objects, such as pulsars, and particular neutron
stars. Some examples include the so-called magnetars and strange quark
magnetars. In particular, NLED modifies in a fundamental basis the
concept of Gravitational Redshift as compared to the well-established
method introduced by standard general relativity \cite{key-7}. The
analysis in Ref. \cite{key-7} proved that unlike general relativity,
where the Gravitational Redshift is independent of any background
magnetic field, when a NLED is incorporated into the photon dynamics,
an effective Gravitational Redshift appears, which happens to depend
decidedly on the magnetic field pervading the pulsar. An analogous
result has also been obtained in Ref. \cite{key-8} for magnetars
and strange quark magnetars. The resulting Gravitational Redshift
tends to infinity as the magnetic field grows larger \cite{key-7,key-8},
as opposed to the predictions of standard general relativity. 

What it is important for our goal in this letter is that the Gravitational
Redshift of neutron stars is connected to the mass\textendash{}radius
relation of the object, see Refs. \cite{key-7} and \cite{key-8}.
Thus, NLED effects turn out to be important as regard to the mass\textendash{}radius
relation, and one can also reasonably expect important effects in
the case of black holes, where the mass\textendash{}radius ratio is
even more important than for a neutron star. Then, from a physical
point of view, the formal analysis presented in the letter displays
a correct procedure to estimate the crucial physical properties stemming
from NLED effects in the presence of super strong magnetic fields.
In fact, the formal discussion developed here shows that the quintessential
density term permits one to construct a model of star supported against
self-gravity entirely by radiation pressure. 

In this sense, the approach here is similar to the one introduced
in the recent papers of Refs. \cite{key-20} and \cite{key-28}, where
it has been shown that it is possible to have radiation pressure supported
stars at arbitrary value of high mass. The result is that trapped
surfaces as defined in general relativity are not formed during a
gravitational collapse, and hence the singularity theorem on black
holes, as proposed in Ref. \cite{key-5}, cannot be applied. 

It is well known that the conditions concerning the early era of the
universe, when very high values of curvature, temperature and density
were present \cite{key-1}, and where matter should be identified
with a primordial plasma \cite{key-1}, \cite{key-9}-\cite{key-11},
are similar to the conditions concerning black holes physics. This
is exactly the motivation because the singularity theorem on black
holes has been generalized to the Universe \cite{key-1,key-10,key-11}.
In the literature there are various cases where a particular analysis
on black holes has been applied to the Universe \cite{key-1,key-10,key-11}
and vice versa \cite{key-1,key-10,key-11,key-14}. 

We will re-discuss the simple case of a homogeneous and isotropic
sphere that was analysed in the historical paper of Oppenheimer and
Snyder \cite{key-12} and, in a different approach, by Beckerdoff
and Misner \cite{key-13}. See also Ref. \cite{key-1} for a review
of the issue. In such a case, the well-known Robertson\textendash{}Walker
line-element can be used, which represents comoving hyper-spherical
coordinates for the interior of the star \cite{key-1}, and in terms
of conformal time $\eta$ it reads

\begin{equation}
ds^{2}=a(\eta)(-d\eta^{2}+d\chi^{2}+\sin^{2}\chi(d\theta^{2}+\sin^{2}\theta d\varphi^{2}),\label{eq: metrica conformemente piatta}\end{equation}

where $a(\eta)$ is the scale factor of a conformal space-time. Using
$\sin^{2}\chi$ we are choosing the case of positive curvature, which
is the only one of interest because it corresponds to a gas sphere
whose dynamics begins at rest with a finite radius \cite{key-1}. 

In Refs. \cite{key-1,key-12} and \cite{key-13} the simplest model
of a \emph{{}``star of dust''} has been discussed, i.e. the case
of zero pressure. In this case the stress-energy tensor is 

\begin{equation}
T=\rho u\otimes u,\label{eq: stress energy}\end{equation}

where $\rho$ is the density of the star and $u$ the four-vector
velocity of the matter. Thus, working with $8\pi G=1$, $c=1$ and
$\hbar=1$ (natural units), following Ref. \cite{key-9}, the Einstein
field equations give only one meaningful relation

\begin{equation}
(\frac{da}{d\eta})^{2}+a^{2}=\frac{\rho}{3}a^{4},\label{eq: Einstein 1}\end{equation}

which admits the familiar cycloidal solution 

\begin{equation}
a=\frac{a_{0}}{2}(1+\cos\eta),\label{eq: cicloide}\end{equation}

where $a_{0}$ is a constant, that is singular for $\eta=\pi.$

Now, let us realize our modified model. We re-introduce a more general
stress-energy tensor, i.e. the one regarding a relativistic perfect
fluid \cite{key-1,key-9}:

\begin{equation}
T=\rho u\otimes u-pg,\label{eq: stress energy 2}\end{equation}

where $p$ is the pressure and $g$ is the metric. In this case, by
restoring the ordinary time with the substitution \cite{key-1,key-9}
\begin{equation}
dt=ad\eta,\label{eq: conforme}\end{equation}

the Einstein field equations give the relation\begin{equation}
(\dot{a})^{2}=\frac{\rho}{3}a^{2}-1,\label{eq: Einstein 2}\end{equation}

where dot represents time-derivative.

We will use the non-linear electrodynamics Lagrangian studied in Ref.
\cite{key-9}. It describes the Heisenberg-Euler NLED

\begin{equation}
\mathcal{L}_{m}\equiv-\frac{1}{4}F+c_{1}F^{2}+c_{2}G^{2},\label{eq: NLD}\end{equation}

where $G=\frac{1}{2}\eta_{\alpha\beta\mu\nu}F^{\alpha\beta}F^{\mu\nu}$,
$F\equiv F_{\mu\nu}F^{\mu\nu}$ is the electromagnetic scalar and
$c_{1}$ and $c_{2}$ are constants \cite{key-9}.

Thus, we also use the equation of state

\begin{equation}
p=\frac{1}{3}\rho-\rho_{\gamma},\label{eq: star}\end{equation}

where (see Eq. (25) of Ref. \cite{key-9}) 

\begin{equation}
\rho_{\gamma}\equiv\frac{16}{3}c_{1}B^{4}.\label{eq: def star}\end{equation}
$B$ is the strength of the magnetic field associated to $F$. In
Ref. \cite{key-9} the equation of state Eq. (\ref{eq: star}) has
been obtained by performing an averaging on electric and magnetic
fields. The presence of the \emph{quintessential density term} $\rho_{\gamma}$
will permit to violate the Null-Energy-Condition of the Penrose's
Theorem \cite{key-5}.

The equation of state (\ref{eq: star}) is no longer given by the
Maxwellian value. Thus, by inserting Eq. (\ref{eq: star}) and Eq.
(\ref{eq: def star}) into Eq. (\ref{eq: Einstein 2}) one obtains
\cite{key-9}\begin{equation}
\dot{a}^{2}=\frac{B_{0}^{2}}{6a^{2}}\left(1-\frac{8c_{1}B_{0}^{2}}{a^{4}}\right)-1,\label{eq: Hubble 2}\end{equation}

which can be solved as obtained by \cite{key-9}

\begin{equation}
t=\int_{a(0)}^{a(t)}dz(\frac{B_{0}^{2}}{6z^{2}}-\frac{8c_{1}B_{0}^{4}}{6z^{6}}-1)^{-\frac{1}{2}}.\label{eq: soluzione}\end{equation}

This expression is not singular for values of $c_{1}>0$. In fact,
by using elliptic functions of the first and of second kinds, one
gets a parabolic trend near a minimum value in $t=t_{f}$ for $a(t)$,
where $t_{f}$ is the time which corresponds to the conformal time
$\eta=\pi.$

In concrete terms, by calling $l,m,n$ the solutions of the equation
$8c_{1}B_{0}^{4}-B_{0}^{2}x+3x^{3}=0,$ it is \cite{key-9}

\begin{equation}
\begin{array}{c}
t=[-(m-l)^{\frac{1}{2}}A(\arcsin\sqrt{\frac{z-l}{m-l}},\sqrt{\frac{l-m}{l-n}})\\
\\+n(m-l)^{-\frac{1}{2}}B(\arcsin\sqrt{\frac{z-l}{m-l}},\sqrt{\frac{l-m}{l-n}})]|_{z=a^{2}(0)}^{z=a^{2}(t)},\end{array}\label{eq: soluzione-1}\end{equation}

where \begin{equation}
A(x,y)\equiv\int_{0}^{\sin x}dz[(1-z^{2})^{-1}(1-y^{2}z^{2})^{-1}]\label{eq: ell1}\end{equation}
 is the elliptic function of the first kind and \begin{equation}
B(x,y)\equiv\int_{0}^{\sin x}dz[((1-z^{2})^{-1})^{-\frac{1}{2}}((1-y^{2}z^{2})^{-1})^{\frac{1}{2}}]\label{eq: ell2}\end{equation}
 is the elliptic function of the second kind.

Then, by defining the minimum value of the scale factor like 

\begin{equation}
a_{f}\equiv a(t=t_{f}),\label{eq: a min}\end{equation}

and recalling that the Schwarzschild radial coordinate, in the case
of the black hole geometry (\ref{eq: metrica conformemente piatta}),
is \cite{key-1}

\begin{equation}
r=a\sin\chi_{0},\label{eq: mach}\end{equation}

where $\chi_{0}$ is radius of the surface in the coordinates (\ref{eq: metrica conformemente piatta}),
one gets

$r_{f}=a_{f}\sin\chi_{0}>2M$ if $B_{0}$ has an high strength, where
$M$ is the black hole mass and $2M$ the gravitational radius in
natural units \cite{key-1}. Thus, we find that the mass of the star
generates a curved space-time without event horizons. 

Now, let us consider the famous Oppenheimer - Volkoff Equation \cite{key-15}
($r$ the radial coordinate)

\begin{equation}
\frac{dp}{dr}=-\frac{(M+4\pi r^{3}p)(p+\rho)}{r(r-2M)}.\label{eq: Oppenheimer - Volkoff}\end{equation}

Using Eq. (\ref{eq: star}) one gets

\begin{equation}
\rho_{f}=\rho_{0}-\intop_{r_{0}}^{r_{f}}\frac{[3M+4\pi r^{3}(\rho-3\rho_{\gamma})](4\rho-3\rho_{\gamma})}{r(r-2M)}dr,\label{eq: o-v 2}\end{equation}

where $r_{f}\equiv r(t_{f})$, $r_{0}\equiv r(0)$, $\rho_{f}\equiv\rho(t_{f})$,
$\rho_{0}\equiv\rho(0)$.

By defining 

\begin{equation}
\begin{array}{ccc}
\rho_{f}\equiv\rho_{f1}+\rho_{f2}+\rho_{f3}+\rho_{f4}+\rho_{f5}, &  & \rho_{f1}\equiv\rho_{01}-\intop_{r_{0}}^{r_{f}}\frac{12\rho M}{r(r-2M)}dr,\\
\\\rho_{f2}\equiv\rho_{02}+\intop_{r_{0}}^{r_{f}}\frac{9\rho_{\gamma}M}{r(r-2M)}dr, &  & \rho_{f3}\equiv\rho_{03}-\intop_{r_{0}}^{r_{f}}\frac{16\pi r^{3}\rho^{2}}{r(r-2M)}dr,\\
\\\rho_{f4}\equiv\rho_{04}+\intop_{r_{0}}^{r_{f}}\frac{60\pi r^{3}\rho\rho_{\gamma}}{r(r-2M)}dr, &  & \rho_{f5}\equiv\rho_{05}-\intop_{r_{0}}^{r_{f}}\frac{36\pi r^{3}\rho_{\gamma}^{2}}{r(r-2M)}dr,\end{array}\label{eq: decomposizione}\end{equation}

the integral (\ref{eq: o-v 2}) can be computed by separating variables.

One then gets

\begin{equation}
\begin{array}{c}
\rho_{f}=\rho_{01}\frac{r_{f}}{r_{0}}\frac{r_{0}-2M}{r_{f}-2M}+\rho_{02}+\frac{9}{2}\rho_{\gamma}\ln\frac{r_{0}}{r_{f}}\frac{r_{f}-2M}{r_{0}-2M}+\\
\\+\{\rho_{03}^{-1}16\pi[r_{0}(r_{0}-r_{f})+\frac{1}{2}(r_{0}^{2}-r_{f}^{2})+4M^{2}\ln\frac{r_{f}-2M}{r_{0}-2M}]\}^{-1}+\\
\\+\rho_{04}\exp60\pi[r_{0}(r_{0}-r_{f})+\frac{1}{2}(r_{0}^{2}-r_{f}^{2})+4M^{2}\ln\frac{r_{f}-2M}{r_{0}-2M}]+\\
\\+\rho_{05}+36\pi\rho_{\gamma}^{2}[r_{0}(r_{0}-r_{f})+\frac{1}{2}(r_{0}^{2}-r_{f}^{2})+4M^{2}\ln\frac{r_{f}-2M}{r_{0}-2M}].\end{array}\label{eq: densita finale}\end{equation}

A reasonable assumption is $\rho\rightarrow0$ for $r\rightarrow r_{0,}$
i.e. at the beginning of the collapse the initial density is considered
negligible with respect to the final density. In this case it is $\rho_{01}=\rho_{02}=\rho_{f03}=\rho_{04}=\rho_{05}\simeq0,$
and the final density reduces to 

\begin{equation}
\rho_{f}=\frac{9}{2}\rho_{\gamma}\ln\frac{r_{0}}{r_{f}}\frac{r_{f}-2M}{r_{0}-2M}+36\pi\rho_{\gamma}^{2}[r_{0}(r_{0}-r_{f})+\frac{1}{2}(r_{0}^{2}-r_{f}^{2})+4M^{2}\ln\frac{r_{f}-2M}{r_{0}-2M}].\label{eq: finale v2}\end{equation}

Thus, we have shown that the density singularity has also been removed
and that, fixed the radius of the star, the final density depends
only by the introduced \emph{quintessential density term} $\rho_{\gamma}$
and by the mass.
\begin{itemize}
\item Notice that in this last updated version we correct two typos which
were present in Eqs. (\ref{eq: densita finale}) and (\ref{eq: finale v2})
in the version of this letter which has been published in Mod. Phys.
Lett. A 25, 2423-2429 (2010). In the present version, both of Eqs.
(\ref{eq: densita finale}) and (\ref{eq: finale v2}) are dimensionally
and analytically correct.
\end{itemize}
In summary, adapting to the black hole case a cosmological analysis
in Ref. \cite{key-9}, we have shown that the introduction of the
Heisenberg-Euler NLED (\ref{eq: NLD}) in the black hole physics can,
in principle, remove the black hole singularity for values of $c_{1}>0$
in Eq. (\ref{eq: NLD}). The analysis has been be improved by applying
the Oppenheimer - Volkoff Equation \cite{key-15} to the black hole
case. In this way, the density singularity of a black hole has also
been removed. We found that at the end, fixed the radius of the star,
the final density depends only by $\rho_{\gamma}$ and by the mass
of the black hole.

We are going to further improve the analysis by using the Born-Infeld
Lagrangian in the context of black holes physics in a future work
\cite{key-16}.

Finally, we take the chance to stress that this result follows in
the same direction of the papers on this issue pioneered by Mitra
\cite{key-17,key-21,key-22} and Robertson, Leiter and Schild \cite{key-18},
\cite{key-23}-\cite{key-27}, who have shown that the formation of
trapped surfaces during the gravitational collapse can be avoided
if the collapsing star develops a powerful magnetic field of strength
of $>10^{10}G$, or equivalently, an intrinsic magnetic moment on
the order of $10^{29}G-cm^{3}$, in the case of galactic black hole
candidates. 

For a sake of completeness \cite{key-29}, we recall previous works
attempting to remove black holes singularities using nonlinear electrodynamics
by Ayon-Beato and Garcia in \cite{key-30}-\cite{key-34}.

\subsection*{Acknowledgements}

The authors thank an unknown reviewer for helpful advices which permitted
to improve this work.

\end{document}